\documentclass[sigconf,screen]{acmart}

\copyrightyear{2023}
\acmYear{2023}
\setcopyright{acmlicensed}
\acmConference[ESEC/FSE '23]{Proceedings of the 31st ACM Joint European Software Engineering Conference and Symposium on the Foundations of Software Engineering}{December 3--9, 2023}{San Francisco, CA, USA}
\acmBooktitle{Proceedings of the 31st ACM Joint European Software Engineering Conference and Symposium on the Foundations of Software Engineering (ESEC/FSE '23), December 3--9, 2023, San Francisco, CA, USA}
\acmPrice{15.00}
\acmDOI{10.1145/3611643.3616242}
\acmISBN{979-8-4007-0327-0/23/12}

\usepackage[utf8]{inputenc} 
\usepackage[T1]{fontenc}
\usepackage{microtype}
\usepackage{tabularx}
\usepackage{multirow}
\usepackage{booktabs}
\usepackage{algorithm}
\usepackage[noend]{algpseudocode}
\usepackage{xcolor}
\usepackage{colortbl}

\usepackage{graphicx}
\usepackage{url}
\usepackage{paralist}
\usepackage{multicol}
\usepackage{pbox}
\usepackage{mathtools}
\usepackage{graphicx}
\usepackage{url}
\usepackage{balance}
\usepackage{layout}
\usepackage{amsmath}
\usepackage{setspace}
\usepackage{verbatim}
\usepackage{float}
\usepackage[normalem]{ulem}
\usepackage{tikz}
\usepackage{ifthen}
\usepackage{enumitem}
\usepackage{subcaption}

\def\BibTeX{{\rm B\kern-.05em{\sc i\kern-.025em b}\kern-.08em
    T\kern-.1667em\lower.7ex\hbox{E}\kern-.125emX}}

\usepackage{marginnote}

\usepackage{xspace}

\usepackage{listings}
\usepackage[scaled]{beramono}
\newcommand*\LSTfont{\Small\fontencoding{T1}\ttfamily\SetTracking{encoding=*}{-60}\lsstyle}
\usepackage{balance} %

\usepackage{cleveref}

\begin{document}

\newcommand{\todo}[1]{{\color{red}\bfseries [[#1]]}}

\newboolean{showcomments}
\setboolean{showcomments}{false}

\ifthenelse{\boolean{showcomments}}{
}
{
	\renewcommand{\todo}[1]{\relax}
}

\newif\ifanonymous
\anonymoustrue

\newcommand{\anonurl}[1]{\ifanonymous URL removed for anonymity.\else\url{#1}\fi}
\newcommand{\footnoteanonurl}[1]{\footnote{\anonurl{#1}}}

\def\|#1|{\mathid{#1}}
\newcommand{\mathid}[1]{\ensuremath{\mathit{#1}}}
\def\<#1>{\codeid{#1}}
\protected\def\codeid#1{\ifmmode{\mbox{\sf{#1}}}\else{\sf #1}\fi}

\hyphenation{type-state}        %

\newcommand{\prefigcaption}{\vspace{-5pt}}
\newcommand{\posttablecaption}{\vspace{-5pt}}

\addtolength{\textfloatsep}{-.25\textfloatsep}
\addtolength{\dbltextfloatsep}{-.25\dbltextfloatsep}
\addtolength{\floatsep}{-.25\floatsep}
\addtolength{\dblfloatsep}{-.25\dblfloatsep}

\ifthenelse{\boolean{showcomments}}
{\newcommand{\nb}[2]{
		\fbox{\bfseries\sffamily\scriptsize#1}
		{\sf\small$\blacktriangleright$\textit{#2}$\blacktriangleleft$}
	}
	\newcommand{\cvsversion}{\emph{\scriptsize$-$Id: macro.tex,v 1.9 2005/12/09 22:38:33 giulio Exp $}}
}
{\newcommand{\nb}[2]{}
	\newcommand{\cvsversion}{}
}

\newcommand\os[1]{{\color{red} \nb{OSCAR}{#1}}}
\newcommand\martin[1]{{\color{blue} \nb{MARTIN}{#1}}}
\newcommand\kobi[1]{{\color{blue} \nb{KOBI}{#1}}}

\newcommand\tod[1]{{\color{red} \nb{TODO}{#1}}}
\newcommand\new[1]{{\color{blue} {#1}}}
\newcommand\re{{\color{red} [REFS]}}
\newcommand\rf{\re}
\newcommand\refs{\rf}

\newcommand\fix[1]{{\color{blue} \nb{FIX THIS}{#1}}}
\newcommand{\here}{{\color{blue} \nb{***}{continue}}}

\newcommand{\ie}{\textit{i.e.},\xspace}
\newcommand{\eg}{\textit{e.g.},\xspace}
\newcommand{\etc}{\textit{etc.}\xspace}
\newcommand{\etal}{\textit{et al.}\xspace}
\newcommand{\aka}{\textit{a.k.a.}\xspace}	

\newlist{researchquestions}{enumerate}{1}
\setlist[researchquestions]{label*=\textbf{RQ\arabic*}}

\newcommand\rev[1]{#1}

\title{On the Relationship {b}etween Code Verifiability and Understandability}

\author{Kobi Feldman}
\email{jcfeldman@wm.edu}
\affiliation{
  \institution{College of William \& Mary}
  \city{Williamsburg}
  \state{Virginia}
  \country{USA}
}

\author{Martin Kellogg}
\email{martin.kellogg@njit.edu}
\affiliation{
  \institution{New Jersey Institute of Technology}
  \city{Newark}
  \state{New Jersey}
  \country{USA}
}

\author{Oscar Chaparro}
\email{oscarch@wm.edu}
\affiliation{
  \institution{College of William \& Mary}
  \city{Williamsburg}
  \state{Virginia}
  \country{USA}
}

\begin{abstract}
  Proponents of software verification have argued that simpler code is easier
  to verify: that is, that verification tools issue fewer false positives and require
  less human intervention when analyzing simpler code. We
  empirically validate this assumption by comparing the number
  of warnings produced by
  four state-of-the-art verification tools
  on 211 snippets of Java code
  with 20 metrics of code comprehensibility
  from human subjects in six prior studies.

  Our experiments, based on a statistical (meta-)analysis, show that, in aggregate, there is a small correlation ($r=0.23$)
  between understandability and verifiability. 
  The results support the
  claim that easy-to-verify code is often easier to understand than
  code that requires more effort to verify.
  Our work has implications for
  the users and designers of verification tools and for future
  attempts to automatically measure code comprehensibility:
  verification tools may have ancillary benefits to
  understandability, and measuring understandability %
  may require reasoning about semantic, not just syntactic, code properties.
\end{abstract}




\maketitle

\section{Introduction}
\label{sec:intro}
Programmers must deeply understand source code in order to
implement new
features, fix bugs, refactor, review code, and do other essential
software engineering
activities~\cite{Tao:FSE12,Peitek:ICSE21,Ammerlaan:SANER15,Maalej:TOSEM14}.
However, understanding code is %
challenging and
time-consuming for developers:
studies~\cite{Xia:TSE18,Minelli:ICPC15} have
estimated developers spend 58\%--70\% of their time
understanding code.

Complexity is a major reason why code can be hard to understand~\cite{Ajami:EMSE19,Scalabrino:TSE19,Peitek:ICSE21,Antinyan:EMSE17,Antinyan:IEEE20}:
algorithms may be written in convoluted ways or be composed of
numerous interacting code structures and dependencies.
There are two major sources of complexity in code:
\emph{essential complexity}, which is needed for the code
to work, and \emph{accidental complexity}, which could be
removed while retaining the code's semantics~\cite{Antinyan:IEEE20,brooks1987no}.
Whether the complexity is essential
or accidental, understanding complex code demands high cognitive effort
from   developers~\cite{Scalabrino:TSE19,Ajami:EMSE19}.

Researchers have proposed many metrics to approximate code
complexity~\cite{Nunez-Varela:JSS17,Curtis:TSE79,Zuse:IWCP'93,Sneed:JSMRP95,Ajami:EMSE19,Jbara:EMSE17,Chidamber:TSE94,Henderson-Sellers1995}
using
vocabulary size (\eg Halstead's complexity~\cite{Halstead1977}),
program execution paths (\eg McCabe's cyclomatic complexity~\cite{McCabe:TSE76}),
program data flow (\eg Beyer's DepDegree~\cite{Beyer:ICPC10}), \etc
These syntactic metrics are intended to
alert developers about complex code so they can refactor
or simplify it to remove accidental complexity~\cite{Garcia-Munoz2016,Peitek:ICSE21,Ammerlaan:SANER15}, or to
predict developers' cognitive
load when understanding code
\cite{Peitek:ICSE21,MunozBaron:ESEM20,Scalabrino:TSE19}. However,
recent studies have found that (some of)
  these metrics (\eg McCabe's) either weakly or do not
correlate at all with code understandability as perceived by
developers or measured by their behavior and brain
activity~\cite{Scalabrino:TSE19,Peitek:ICSE21,Feigenspan:ESEM11}.
Other studies have demonstrated that certain code
structures (\eg if \textit{vs} for loops, flat \textit{vs} nested constructs, or
repetitive code patterns) lead to higher or lower understanding effort (\aka \textit{code understandability} or \textit{comprehensibility})~\cite{Ajami:EMSE19,Jbara:EMSE17,Johnson:ICSME19,Feigenspan:ESEM11,Langhout:ICPC21,Borstler:TSE16}, which
diverges from the simplistic way metrics (\eg McCabe's) measure code
complexity~\cite{Ajami:EMSE19,Jbara:EMSE17,Scalabrino:TSE19,Kaner2004,Feigenspan:ESEM11}.

In this paper, we investigate the relationship between %
\rev{understandability} and
\textit{code verifiability}---how easy or hard it is for \rev{a developer to use} a verification
tool to prove safety properties about the code, such as the absence of
null pointer violations or out-of-bounds array accesses.
Our research is motivated by the common assumption in the software
verification community that \emph{simpler code is both easier to verify by verification tools and easier to understand
  by developers}.  For example, the Checker
Framework~\cite{PapiACPE2008} user manual states this assumption
explicitly in its advice about
unexpected warnings: ``rewrite your code to be simpler for
the checker to analyze; this is likely to make it easier for people to
understand, too''~\cite{cfmanual-warnings}. The documentation of the
OpenJML verification tool says~\cite{openjml}: ``success in checking the consistency of
the specifications and the code will depend on... 
the complexity and style in which the code and specifications are
written''~\cite{openjml-manual-complexity}.
This assumption is widely held by verification experts but has never been validated empirically. \looseness=-1

The intuition behind this assumption is that a verifier
can handle a certain amount of code complexity before it issues
a warning. If it is possible to remove
the warning by changing the code, then the complexity that
caused it must be accidental rather than essential,
and therefore removing the warning reduces the
overall complexity of the code.
For example, consider accessing a possibly-null pointer
in a Java-like language.
A simple null check might use an \<if> statement.
A more complex variant with the same semantics might
dereference the
pointer within a \<try> statement and use a \<catch>
statement to intercept the resulting exception if the
pointer is \<null>.
The second,
more convoluted variant (with its significant accidental complexity)
might not be verified---a null pointer dereference does occur, but it is intercepted
before it crashes the program. A verifier would need to
model exceptional control flow to avoid a false positive warning. 
\rev{
Alternatively, a verifier might warn about code that
makes unstated assumptions. For example, by dereferencing a
possibly-null pointer without checking it first, the code assumes
that the pointer has already been checked. A verifier might
warn that this is unsafe unless a human provides a specification
that the pointer is non-null. In that case, the verifier can
verify the dereference, but then must check that the value assigned
to the variable really is non-null at each assignment.
A warning because of a missing specification can also indicate
complexity that a human might need to reason about to understand the code:
the human might need to determine why it is safe to dereference the pointer. 
}

Our goal is to empirically validate the purported relationship
between verifiability and understandability---and
therefore either confirm or refute the assumption that easy-to-understand
code is easy to verify (and vice-versa).
To do so, we need to measure verifiability.
Verifiers analyze source code to
prove the absence of particular classes of defects
(\eg null dereferences) using \emph{sound}
analyses. A sound verifier can find all defects
(of a well-defined class) in
the code. However, most interesting properties of programs
are undecidable~\cite{rice1953classes}, so
all sound verifiers produce false positive warnings: that is,
they
conservatively issue a
warning when they cannot produce a proof.
The user of the verifier must sort the true positive warnings
that correspond to
real bugs from false positive warnings due to the
verifier's imprecision \rev{or due to the need to state code assumptions as specifications}.
The \rev{combination of false positive warnings and warnings about unstated assumptions (which we refer to as  ``false positives'' or ``warnings'' for brevity)} is a good proxy
for verifiability because the fewer such warnings %
in a given piece of code, the less work a developer using the
verifier will need to do to verify that code.
\looseness=-1
With that in mind, \emph{we \textbf{hypothesize} that a correlation exists
between a code snippet's comprehensibility, as judged by humans,
and its verifiability, as measured by false positive warnings}.

\begin{comment}
We hypothesize that the reason that proponents of verification suggest
``making code simpler'' as a possible solution when encountering a
false positive warning could be that \emph{verification tools model
  how humans understand code}: that is, that there is a correlation
between the comprehensibility of a snippet of code as judged by humans
and its verifiability.
\end{comment}

\begin{comment}
If the code
fits within that reasoning framework (assuming the code has no bugs), the verifier likely can verify it. Otherwise, either a false positive warning might occur or a human needs to write a specification of some kind, either in the code under analysis or in one of its dependencies.

"Our intuition here is that program verifiers kind of reason like a human (humans wrote the rules they use to verify programs!), but can't handle edge cases or unexpected code patterns well - so if the verifier can figure it out, it should be easy for a human, too."
\end{comment}

%
%
We conducted an empirical study to validate this hypothesis---the first time that this
common assumption in the verification community is tested empirically.
Our study compares the number
of warnings produced by three state-of-the-art, sound
static code verifiers~\cite{PapiACPE2008,jatyc,openjml} and one industrial-strength, unsound
static analysis tool based on a sound core~\cite{infer} with
\rev{$\approx$18k} measurements code understandability proxies collected from humans in
six prior
studies~\cite{Siegmund:ICSE14,Peitek:TSE18,Raymond:TSE10,Scalabrino:TSE19,Borstler:TSE16,Peitek:ICSE21}
for 211 Java code snippets. Such measurements come from 20 metrics
in four categories~\cite{MunozBaron:ESEM20}: (1)
human-judged \emph{ratings}, (2) program output \emph{correctness},
(3) comprehension \emph{time}, and (4)  \emph{physiological} (\ie brain
activity) metrics.
We used a statistical meta-analysis technique~\cite{Borenstein2009,harrer2021doing} to examine
the correlation between verifiability and these understandability
metrics in aggregate. Given the small sample sizes of the original
studies and the danger of multiple comparisons, this meta-analysis
technique permits us to draw methodologically-sound conclusions
about the overall trends. %

We found %
a small correlation between verifiability and the proxies for understandability, in aggregate ($r=0.23$);
individually, 13 of 20 metrics were correlated with verifiability.
This trend suggests that more often than not,
code that is easier to verify is easier for humans to understand.
\looseness=-1

One implication of this result is that
our results provide evidence for a
relationship between the \emph{semantics} of a piece of code
and its understandability, which may explain (in part) the apparent
ineffectiveness of prior \emph{syntactic} approaches.
This implies that the verifiability of a code snippet, as measured
automatically by the warnings issued by extant verification tools, might be a useful
input to models of code understandability~\cite{Raymond:TSE10,Scalabrino:TSE19,Trockman:MSR18}. 
Another implication is that
when using a verification tool, developers should \emph{consider}
making changes to the code to make it easier to verify automatically: doing
so is \emph{more likely than not} to make the code easier for a human
to understand. If a developer makes a change to remove a false positive
without changing the code's semantics,
any complexity they remove must have been accidental.
This means that verification tools provide a secondary
benefit beyond their guarantees of the absence of errors: code that can be
easily verified should be easier for future developers to improve and extend.
In summary, the main contributions of this paper are:
\begin{itemize}
\item empirical evidence of a correlation between code understandability
  and verifiability derived via meta-analysis,
  supporting the common assumption that easier-to-verify code 
  is easier for humans to understand (and vice-versa). The
  results have implications for the design and deployment of
  verification tools and for developing more accurate automated metrics of code
  comprehensibility; and
\item an online replication package~\cite{repl-pack} that enables verification and
  replication of our results and future
  research.%
  \os{to update the citation of replication package and check it is fixed everywhere} 
\end{itemize}

\section{Empirical Study Design}
\label{sec:methodology}

Our goal is to assess the correlation between
human-based code comprehensibility metrics and code verifiability---\ie
how many warnings static code verification tools issue.
Intuitively, our goal is to check if code that is
easy to verify is also easy for humans to understand.
To that end, we formulate these research questions~(RQs):

\begin{researchquestions}
\item How do \textbf{individual} human-based code comprehensibility metrics
  correlate with tool-based code verifiability?

\item How do human-based code comprehensibility metrics correlate with tool-based
  code verifiability \textbf{in aggregate}?
  
\item What is the impact of each verification tool on the aggregated correlation results?

\item Do different kinds of comprehensibility metrics correlate better or worse with tool-based verifiability?
  
\end{researchquestions}
\begin{table*}[t]
  \setlength\tabcolsep{2.5pt}
  \centering
  \caption{Datasets (DSs) of code snippets and understandability measurements/metrics used in our study.
  The metrics types are ``C'' for correctness, ``R'' for ratings, ``T'' for time, and ``P'' for physiological.}
  \label{tab:datasets}
  \resizebox{\textwidth}{!}{%
  \begin{tabular}{cccp{20mm}llr}
    \hline
  \textbf{DS} & \textbf{Snippets} & \textbf{NCLOC} & \textbf{Participants} & \textbf{Understandability   Task} & \multicolumn{1}{c}{\textbf{Understandability Metrics}} & \textbf{Meas.} \\ \hline
  1~\cite{Siegmund:ICSE14} & 23   CS algorithms & 6   - 20 & 41   students & Determine   prog. output & \begin{tabular}[c]{@{}l@{}}\textbf{C:} \textit{correct\_output\_rating}   (3-level correctness score for program output)\\      \textbf{R:} \textit{output\_difficulty} (5-level difficulty score for program   output)\\      \textbf{T:} \textit{time\_to\_give\_output} (seconds to read program and answer a question)\end{tabular} & 2,829 \\ \hline
  2~\cite{Peitek:TSE18} & 12   CS algorithms & 7   - 15 & 16   students & Determine   prog. output & \begin{tabular}[c]{@{}l@{}}\textbf{P:} \textit{brain\_deact\_31ant}   (deactivation of brain area BA31ant)\\      \textbf{P:} \textit{brain\_deact\_31post} (deactivation of brain area BA31post)\\      \textbf{P:} \textit{brain\_deact\_32} (deactivation of brain area BA32)\\      \textbf{T:} \textit{time\_to\_understand} (seconds  to   understand program within 60 seconds)\end{tabular} & 228 \\ \hline
  3~\cite{Raymond:TSE10} & 100   OSS methods & 5   - 13 & 121   students & Rate   prog. readability & \textbf{R:} \textit{readability\_level} (5-level score for readability/ease to   understand) & 12,100 \\ \hline
  6~\cite{Scalabrino:TSE19} & 50   OSS methods & 18   - 75 & 50 students and 13   developers & Rate   underst./answer Qs & \begin{tabular}[c]{@{}l@{}}\textbf{R:} \textit{binary\_understandability} (0/1   program understandability score)\\      \textbf{C:} \textit{correct\_verif\_questions} (\% of correct answers to verification   questions)\\      \textbf{T:} \textit{time\_to\_understand} (seconds  to   understand program)\end{tabular} & 1,197 \\ \hline
  9~\cite{Borstler:TSE16} & 10   OSS methods & 10   - 34 & 104   students & Rate   read./complete prog. & \begin{tabular}[c]{@{}l@{}}\textbf{C:} \textit{gap\_accuracy} ([0-1] accuracy   score for filling in program blanks)\\      \textbf{R:} \textit{readability\_level\_ba} (5-level avg. score for readability b/a   code completion)\\     \textbf{R:} \textit{readability\_level\_before} (5-level score for readability before code   completion)\\      \textbf{T:} \textit{time\_to\_read\_complete} (avg. seconds to rate readability and complete   code)\end{tabular} & \rev{716} \\ \hline
  F~\cite{Peitek:ICSE21} & 16   CS algorithms & 7   - 19 & 19   students & Determine   prog. output & \begin{tabular}[c]{@{}l@{}}\textbf{P:} \textit{brain\_deact\_31}(deactivation   of brain area BA31)\\      \textbf{P:} \textit{brain\_deact\_32} (deactivation of brain area BA32)\\      \textbf{R:} \textit{complexity\_level} (score for program complexity)\\      \textbf{C:} \textit{perc\_correct\_output} (\% of subjects who correctly gave program output)\\      \textbf{T:} \textit{time\_to\_understand} (seconds  to   understand program within 60 seconds)\end{tabular} & \rev{935} \\
  \hline
  \end{tabular}%
  }
  \vspace{-.2cm}
  \end{table*}

\textbf{RQ1} and \textbf{RQ2} encode our \textit{hypothesis}:
that a correlation exists between tool-based code verifiability and
human-based code comprehensibility.
\textbf{RQ1} asks whether individual metrics correlate with verifiability.
However, due to the limitations of prior studies, sample sizes for the
metrics considered individually are too small to draw
reliable conclusions. Therefore, \textbf{RQ2} asks whether
there is a pattern to the answers to \textbf{RQ1} that summarizes the overall correlation trend.
We answer \textbf{RQ2} statistically by combining the results of the individual
metrics targeted by \textbf{RQ1} with a meta-analysis.

\textbf{RQ3} asks whether verifiability, measured using only
one tool's warnings,
correlates with comprehensibility. Intuitively, we aim to
check (1) if the patterns of correlations are similar across tools,
(indicating generalizability),
and (2) whether any particular tool dominates the results.
To answer the second part, we use a ``leave-one-out''
ablation analysis, dropping each tool individually.
\textbf{RQ4} asks whether there is any difference in correlation
between verifiability and different proxies for code
comprehensibility. Based on prior work~\cite{MunozBaron:ESEM20}, we
focus on four metric categories:
\textit{correctness}, \textit{rating}, \textit{time}, and
\textit{physiological}.
Together, the answers to \textbf{RQ3} and \textbf{RQ4} help us
explain our \textbf{RQ1} and \textbf{RQ2} results: they explore the
tool(s) and metric(s) 
responsible for the observed correlations.\looseness=-1

To answer our RQs, we first compiled a set of human-based code
comprehensibility measurements from prior
studies (\cref{sec:complexity-studies}). Then, we
\rev{defined our metric for code verifiability (\cref{sec:verifiability}) and}
executed four verification-based tools on the same
code snippets to measure how often each snippet cannot be verified
(\cref{sec:verifiers,sec:tool-execution}).
\rev{Next, we conducted an analysis of the warnings produced by the
verifiers to ensure that they met our definition of ``false positive''
(\cref{sec:warning-validation}).}
Finally, we correlated the
comprehensibility metrics with the number of warnings produced by
the tools and analyzed the
correlation results using a meta-analysis (\cref{sec:correlation}). 

We did a correlation study rather than try to establish
causation because that would require expensive controlled experiments
with human subjects. Since correlation cannot exist without causation,
it is practical to re-use existing studies and establish correlation first
before attempting a causation study, which we leave as future work.
\looseness=-1

\vspace{-0.2cm}
\subsection{Code and Understandability Datasets}
\label{sec:complexity-studies}

We used existing datasets (DSs) from six prior understandability studies~\cite{Siegmund:ICSE14,Peitek:TSE18,Raymond:TSE10,Scalabrino:TSE19,Borstler:TSE16,Peitek:ICSE21}, which are summarized in \cref{tab:datasets}. Each
study used a different set of code snippets and proxy metrics to
measure understandability using different groups of human subjects
who performed specific understandability tasks---see \cref{tab:datasets}.

\rev{To select these datasets, we leveraged the systematic literature review conducted by Muñoz
\etal~\cite{MunozBaron:ESEM20}, who found ten studies that measured code understandability with publicly available data. From these ten studies, we selected the five studies whose snippets were written in Java, since the verifiers we consider only work on Java code---see \cref{sec:verifiers}.
To identify the datasets and facilitate
replication, we use the same nomenclature as Muñoz
\etal's: DS1, DS2, DS3, DS6, and DS9. Since those five studies were conducted before 2020, we performed a literature search of additional comprehensibility studies from 2020 to early 2023 and found the one by Peitek \etal~\cite{Peitek:ICSE21} (Dataset F or DSF), who also used Java snippets.
}

In total, we used \rev{$\approx$18k} understandability measurements (see the ``Meas.'' column in \cref{tab:datasets}) for 211 Java code snippets, collected from 364 human subjects using 20 metrics.
The 211 snippet programs are 5 to 75 non-comment/blank LOC or
NCLOC---17 NCLOC on avg.---with different complexity
levels\rev{, as reported by the methodology of their respective studies~\cite{Siegmund:ICSE14,Peitek:TSE18,Raymond:TSE10,Scalabrino:TSE19,Borstler:TSE16,Peitek:ICSE21}}.
Datasets 3, 6, and 9 derive from open source software projects
(OSS)---\eg Hibernate, JFreeChart, Antlr,
Spring, \& Weka~\cite{Raymond:TSE10,Scalabrino:TSE19,Borstler:TSE16}---and
the other datasets are implementations of
algorithms from 1st-year programming courses
(\eg reversing an array)~\cite{Siegmund:ICSE14,Peitek:TSE18,Peitek:ICSE21}.
The original studies selected short code snippets 
to control for potential cofounding factors that may affect
understandability~\cite{Siegmund:ICSE14,Peitek:TSE18,Peitek:ICSE21}.
\looseness=-1

We selected the understandability metrics
used in the meta-study conducted by Muñoz
\etal~\cite{MunozBaron:ESEM20}---see \cref{tab:datasets} for the metrics, their type, and a brief description of them (our replication package has full descriptions~\cite{repl-pack}).
We also used Muñoz \etal's categorization of the metrics.
\textit{Correctness} metrics (marked with a \textbf{C} in \cref{tab:datasets})
measure the correctness of the program output given by the participants.
\textit{Time} (\textbf{T}) metrics measure the time that participants took to read, understand, or complete a snippet.
\textit{Rating} (\textbf{R}) metrics indicate the subjective rating given by the participants about their understanding of the code snippet or code readability, using Likert scales.
\textit{Physiological} (\textbf{P}) metrics measure the concentration level of the participants during program understanding, via deactivation measurements of brain areas (\eg Brodmann Area 31 or BA31~\cite{Siegmund:ICSE14}).

Study participants were mostly CS undergraduate/graduate students with intermediate-to-high programming experience, as reported in the original papers~\cite{Siegmund:ICSE14,Peitek:TSE18,Raymond:TSE10,Scalabrino:TSE19,Borstler:TSE16,Peitek:ICSE21}. Only DS6's study included professional developers~\cite{Scalabrino:TSE19}---see \cref{tab:datasets}. 

\rev{We used all the available data from the six original studies. 
Their measurements come in aggregate or individual form:
\eg the physiological measurements in DS2~\cite{Peitek:TSE18} and DSF~\cite{Peitek:ICSE21} 
are provided per snippet averaged across participants, 
while the DS3 measurements come for each participant and snippet~\cite{Raymond:TSE10}. 
Some studies also included an uneven numbers of participants per snippet, 
due to different methodological decisions. 
For example, DS9's study included a random assignment of participants to one of six sequences of five snippets~\cite{Borstler:TSE16}. 
In DS6's study~\cite{Scalabrino:TSE19}, 
six snippets were understood by eight participants and the remaining 44 snippets were understood by nine. 
For these reasons, each dataset's number of measurements 
(see the ``Meas.'' column in \cref{tab:datasets}) 
is not always divisible by the number of snippets and participants.}

\subsection{Proxy for Code Verifiability}
\label{sec:verifiability}

\rev{We define \textit{code verifiability} as the \textit{effort} that a developer incurs when using a verification
tool to prove safety properties about a snippet of code. Since measuring this effort is infeasible without running a study with verification tool users, we use \textit{false positive counts} as an automatable proxy for verifiability.}

\rev{We define a ``false positive'' as a verifier warning that indicates the verifier is unable to prove that a code snippet is correct due to a weakness in the verifier (\ie undecidablilty~\cite{rice1953classes}) or due to a missing code specification, which a human should provide. In effect, a false positive warning represents a ``fact'' about the code that the verifier needs, but cannot prove with the snippet's code only.
\looseness=-1}

\rev{Our definition of ``false positive'' differs from
  the typical one when evaluating the precision and recall of a verifier.
  In that context, it is assumed that correct specifications are explicit and available, and
  ``false positive'' means a fact that the verifier cannot prove, even with a specification.
  Conversely, in our context, we want a proxy for the \emph{difficulty} of verifying
  a snippet. That difficulty 
  includes both writing specifications and suppressing false positive warnings, so it
  is sensible to include both in our proxy for verifiability. In other words, the fewer warnings a verifier issues, the less work a developer using the verifier must do to verify a code snippet.}

\rev{
We considered two other proxies for verifiability: number of false positives
after writing specifications and number of ``facts'' verified about the code. We discarded the former proxy 
because the full context of how the snippets are intended to be used is not available and because writing specifications is error- and bias-prone. The latter was discarded because none of the verifiers provide the proxy directly,
and because approximating whether or not a verifier needs to even check a fact is undecidable for
some properties considered by a verifier (\eg determining what is considered a resource in the code by a resource leak verifier~\cite{KelloggSSE2021}), so a precise count is impossible.}
\looseness=-1

\begin{table*}
  \caption{Number of snippets each tool warns on and the total number
    of warnings per dataset.
  }
\label{tab:tool-warnings}
\centering
\begin{tabular}{lccccccc|ccccccc}
  \hline
  & \multicolumn{7}{c}{\textbf{Snippets Warned On}} & \multicolumn{7}{c}{\textbf{Total Warnings}} \\\cline{3-7}\cline{10-14}
  \textbf{Tool\: \textbackslash \: Dataset } & \textbf{1} & \textbf{2} & \textbf{3} & \textbf{6} & \textbf{9} & \textbf{F} & \textbf{All} & \textbf{1} & \textbf{2} & \textbf{3} & \textbf{6} & \textbf{9} & \textbf{F} & \textbf{All} \\
  \hline
  \textbf{Infer} & 0/23 & 0/12 & \rev{1/100} & \rev{5/50} & \rev{0/10} & 1/16 & \rev{\textbf{7/211}} & 0 & 0 & \rev{1} & \rev{7} & 0 & 1 & \rev{\textbf{9}} \\
  \textbf{Checker Fr.} & 3/23 & 0/12 & \rev{18/100} & 28/50 & 4/10 & 3/16 & \textbf{56/211} & 7 & 0 & \rev{51} & 83 & 4 & 3 & \textbf{148} \\
  \textbf{JaTyC} & 3/23 & 1/12 & 88/100 & 40/50 & 10/10 & 2/16 & \textbf{144/211} & 14 & 3 & 327 & 537 & 37 & 6 & \textbf{924} \\
  \textbf{OpenJML} & 14/23 & 6/12 & 69/100 & 41/50 & 10/10 & 13/16 & \textbf{153/211} & 29 & 11 & 808 & 219 & 24 & 29 & \textbf{1,120} \\
  \hline
  \textbf{All Tools}  & 17/23 & 7/12 & \rev{93/100} & 48/50 & 10/10 & 15/16 & \rev{\textbf{190/211}} & 50 & 14 & \rev{1,187} & \rev{846} & \rev{65} & 39 & \rev{\textbf{2,201}} \\
  \hline
\end{tabular}
\end{table*}

\subsection{Verification Tools}
\label{sec:verifiers}

We used the following criteria to select verification tools:
\begin{enumerate}
\item Each tool must be based on a sound core---\ie the
  underlying technique must generate a proof.
\item Each tool must be actively maintained.
\item Each tool must fail to verify at least one
  snippet.
\item Each tool must run mostly automatically.
\item Each tool must target Java.
\end{enumerate}

Criterion 1 requires that each tool be verification-based.
Our hypothesis implies that the \emph{process of verification}
can expose code complexity: that is,
our purpose in running verifiers is not to expose
bugs in the code but to observe when the tools produce false positive warnings (due to code complexity).
Therefore, each tool must perform verification under the hood (\ie
must attempt to construct a proof) for our results to be meaningful.
This criterion excludes non-verification static analysis tools such as
FindBugs~\cite{AyewahHMPP2008} which use
unsound heuristics.
Exploring whether those tools correlate
with comprehensibility is future work.
However, criterion~1 does \emph{not} require the tool to be sound: merely that
it be based on a sound core. We permit
\emph{soundiness}~\cite{soundiness-manifesto} (and intentionally-unsound tools) because practical verification
tools commonly only make guarantees about the absence of defects under
certain conditions.

Criteria 2 through 5 are practical concerns.
Criterion 2 requires the verifier to be state-of-the-art
so that our results are useful to the community.
Criterion~3
requires each verifier to issue at least one
warning---for tools that verify a property that is irrelevant
to the snippets (and so do not issue any warnings), we cannot
do a correlation analysis.
Criterion 4 excludes proof assistants and
other tools that require extensive manual effort.
Criterion 5 restricts
the scope of the study: we focused on Java code and verifiers.
We made this choice because (1) 
verifiers are usually language-dependent, 
(2) many prior code comprehensibility studies on
human subjects used Java---\eg 5/10 studies in Muñoz \etal
\cite{MunozBaron:ESEM20} and no other
language has more than 2/10 studies---and (3) Java has received significant
attention from the program verification community due to its
prevalence in practice. We discuss
the threats to validity that this and other choices cause in \cref{sec:threats}.

\subsubsection{Selected Verification Tools}

By applying the criteria defined above, we selected four verification tools: 

\textbf{Infer}~\cite{infer} is
an unsound, industrial static analysis tool based
on a sound core of separation logic~\cite{ohearn2001local}
and bi-abduction~\cite{calcagno2009compositional}.
Separation logic enables reasoning about mutations to program state
independently, making it scalable; bi-abduction
is an inference procedure that automates separation logic reasoning.
Infer is unsound by design: despite internally using a sound,
separation-logic-based analysis, it uses heuristics
to prune all but the most likely bugs from its output, because it
is tailored for deployment in industrial settings.
\rev{Infer warns about possible null dereferences, data races, and resource leaks.}
We used Infer version 1.1.0.
\looseness=-1

The \textbf{Checker Framework}~\cite{PapiACPE2008} is a collection of
pluggable typecheckers~\cite{FosterFFA99}, which
enhance a host language's type system to track
an additional code property, such as whether each object
might be null.
The Checker Framework includes many pluggable typecheckers.
We used the nine that satisfy criterion 4, which prevent programming
mistakes related to:
nullness~\cite{PapiACPE2008,DietlDEMS2011},
interning~\cite{PapiACPE2008,DietlDEMS2011}, object
construction~\cite{KelloggRSSE2020}, resource
leaks~\cite{KelloggSSE2021}, array bounds~\cite{KelloggDME2018}, signature
strings~\cite{DietlDEMS2011}, format strings~\cite{WeitzKSE2014},
regular expressions~\cite{SpishakDE2012}, and
optionals~\cite{optional-checker}. We used Checker Framework
version 3.21.3.
  
The \textbf{Java Typestate Checker (JaTyC)}~\cite{jatyc} is a
typestate analysis~\cite{StromY86}.  A typestate analysis extends a
type system to also track
\emph{states}---for example, a typestate system might track that a
\<File> is first closed, then open, then eventually closed.
Currently-maintained typestate-based Java static analysis tools
include JaTyC~\cite{jatyc} (a typestate verifier) and
RAPID~\cite{emmi2021rapid} (an unsound static analysis tool based on a
sound core that permits false negatives when
verification is expensive). We chose to use JaTyC rather than RAPID
for two reasons. First, JaTyC ships with specifications for general
programming mistakes, but RAPID focuses on mistakes arising from
mis-uses of cloud APIs; the snippets in our study do not interact with
cloud APIs. Second, JaTyC is open-source, but RAPID is closed-source.
\rev{JaTyC warns about possible null dereferences, incomplete protocols
on objects that have a defined lifecycle (such as sockets or files),
and about violations of its ownership discipline, which is similar in 
spirit to Rust's~\cite{klabnik2018rust}.}
We used JaTyC commit b438683.

\textbf{OpenJML}~\cite{openjml}
converts verification conditions to SMT formulae and dispatches those
formulae to an external satisfiability solver.
OpenJML verifies specifications
expressed in the Java Modeling Language (JML)~\cite{leavens1998jml}; it is the
latest in a series of tools verifying JML specifications by reduction
to SMT going back to ESC/Java~\cite{flanagan2002extended}.
\rev{OpenJML verifies the absence of a collection of a common programming errors,
including out-of-bounds array accesses, null pointer dereferences,
integer over- and underflows, and others.}
We used OpenJML 0.17.0-alpha-15 with
the default solver z3~\cite{DeMouraB2008} v. 4.3.1.
\looseness=-1

\subsubsection{Verification Tools Considered but Not Used}
We considered and discarded three other verifiers:
JayHorn~\cite{jayhorn},
which fails criterion 2~\cite{schaef-personal-communication};
CogniCrypt~\cite{KrugerSAKM2018}, which fails criterion 3;
and Java PathFinder~\cite{jpf}, which fails criterion 4.

\subsection{Snippet Preparation and Tool Execution}
\label{sec:tool-execution}

\rev{We acquired the snippets from prior
work~\cite{MunozBaron:ESEM20,Peitek:ICSE21} but had to make some modifications to prepare them 
for tool execution. 
DS3 included 4 commented-out snippets, which we uncommented.
To make the snippets compilable, we 
created ``stubs'' for the classes, method
calls, \etc they use without
modifying the snippets themselves. Since the the snippets themselves did not change, their underlying, measured code comprehensibility did not change either: in the original studies,
the snippets were provided to the humans in isolation. At the same time, our modifications would change the programs' state if the snippets were to be executed. We performed a manual analysis of tool warnings to ensure our modifications did not cause spurious warnings (see \cref{sec:warning-validation}).}
We created scripts to execute the verifiers on
the snippets and display all verification
failures for each tool.
\rev{\Cref{tab:tool-warnings} shows descriptive statistics of the warnings issued by each tool on each dataset.}

\subsection{Code Correctness and Warning Validation}
\label{sec:warning-validation}

\rev{
  We assumed that every warning issued by a verifier about a snippet is a ``false positive'', according to the definition we presented in \cref{sec:verifiability}.
  In effect, this means that we are treating the snippets
  as if they are correct. For example, if a snippet dereferences a pointer without a null
  check, we assume that pointer is non-null; if a snippet accesses an
  array without checking a bound, we assume that the bound was checked elsewhere in
  the program, \etc Each verifier warning therefore represents some fact that the verifier
  needs, but cannot prove with the snippet's code only. We consider these reasonable assumptions because: (1) no context about the snippets is available (or was presented to the human subjects in the prior studies), and (2) the snippets are likely to be correct as researchers showed them to humans in prior comprehensibility studies.

  However, to check that these assumptions do not skew our correlation analysis, we manually validated whether a representative sample of tool warnings were indeed false positives (according to our definition from \cref{sec:verifiability}). 
 After executing the verifiers, one author examined a representative subset of the warnings
  (a sample of 344 of 2,201, at 95\% confidence level and 5\% of error margin) and recorded the cause of each. A second author examined the first author's assessment and both authors discussed the cases where the assessment was incorrect or needed more details, reaching consensus in case of disagreements (of which there were < 5).
  Of these 344 warnings,
  none were ``real bugs'' in the sense that they are guaranteed to make the code fail when executed. Many
  do represent \emph{potential} bugs: that is, code that does not check boundary conditions
  such as nullability; however, these warnings could be removed by writing a specification
  for the relevant verifier indicating the assumptions made by the snippet. This means these warnings are all false positives, according to our definition from \cref{sec:verifiability}. 
  \looseness=-1
  
  In the sample,
  the most common reasons for a verifier to warn were: (1) violation's of JaTyC's Rust-like
  rules for mutability, and 2) violations of the verifiers' assumptions about nullability.
  Other common causes were possible integer over- and underflows, too large or too small array
  indexes, and unsafe casts.
}
Our analysis of warnings for each snippet indicates a fairly uniform
distribution of warning types over the datasets. \rev{Our replication package
provides our detailed analysis of warnings~\cite{repl-pack}.} 

\vspace{-0.2cm}
\subsection{Correlation and Analysis Methods}
\label{sec:correlation}

\subsubsection{Aggregation}
We aggregated the comprehensibility
measurements and the number of tool warnings for each code snippet in
the datasets. The resulting pairs of comprehensibility and
verifiability values per snippet were correlated for
sets of snippets.

Specifically, we averaged the individual code comprehensibility
measurements per snippet for each metric. For example, for each
snippet in DS1 we averaged the 41
\emph{time\_to\_give\_output} measurements collected from the 41
participants in the corresponding study~\cite{Siegmund:ICSE14}.
Following Muñoz \etal~\cite{MunozBaron:ESEM20}, we averaged
discrete measurements, which mostly come from Likert scale responses
in the original studies. For example, the metric
\emph{output\_difficulty} (from DS1) is the perceived
difficulty in determining program output using a 0-4 discrete scale.
While there is no clear
indication of whether Likert scales represent ordinal or continuous
intervals~\cite{Murray2013}, we observed that the Likert items in the original datasets represent  discrete values on continuous scales~\cite{MunozBaron:ESEM20}, so
it is reasonable to average these values to obtain one
measurement per snippet. \rev{All physiological measurements given by the original studies are averaged across all participants who understood a given snippet.}\looseness=-1

Regarding code verifiability, we summed up the number of
warnings from the verification tools for each snippet. 
We considered averaging rather than summing up.
However, since the correlation coefficient that we used
(see below) is robust to data scaling (\ie the average is essentially
a scaled sum), imbalances in the number of warnings from each tool do not change the correlation results. Further, for \textbf{RQ3}, we performed an ablation experiment to investigate possible effects of warning imbalances on correlation.

\subsubsection{Statistical methods}
\label{sec:correlation-methods}
We used Kendall's $\pmb{\tau}$~\cite{Kendall1938} to correlate the 
individual comprehensibility metrics and the tool warnings
because (1) it does not
assume the data to be normally distributed and have a linear
relationship~\cite{Cohen:2002}, (2) it is robust to outliers~\cite{Cohen:2002}, and (3) it has been used
in prior comprehensibility
studies~\cite{Peitek:ICSE21,Scalabrino:TSE19,MunozBaron:ESEM20}. As in
previous studies~\cite{Peitek:ICSE21,Scalabrino:TSE19}, we follow Cohen's guidelines~\cite{Cohen:2002} and
interpret the correlation
strength as \emph{none} when $0 \leq |\tau| < 0.1$,
\emph{small} when $0.1 \leq |\tau| < 0.3$,
\emph{medium} when $0.3 \leq |\tau| < 0.5$,
and \emph{large} when $0.5 \leq |\tau|$.\looseness=-1

To answer \textbf{RQ1}, we first stated the expected correlation (as
either \emph{positive} or \emph{negative}) between each
comprehensibility metric %
and code verifiability that would support our
hypothesis. %
For some metrics, such as \emph{correct\_output\_rating} in DS1, a
\emph{negative} correlation indicates support for the
hypothesis---if humans can deduce the correct output \textit{more}
often, the hypothesis predicts a \emph{lower} number of warnings from
the verifiers. %
A \textit{positive} expected correlation,
such as for \emph{time\_to\_understand} in DS6, indicates that
higher values in that metric support the hypothesis---\eg
if humans take \textit{longer} to understand a snippet, our hypothesis
predicts that \textit{more} warnings will be issued on that snippet.
We computed the correlation (and
its strength) between the comprehensibility metrics and code
verifiability and compared the observed correlations with the expected
ones to check if the results validate or refute our hypothesis.

To answer \textbf{RQ2},
we performed a statistical meta-analysis~\cite{Borenstein2009} of the \textbf{RQ1}
correlation results.
A meta-analysis is
appropriate for answering \textbf{RQ2} because it combines individual
correlation results that come from different metrics as a single aggregated correlation
result~\cite{Borenstein2009,MunozBaron:ESEM20}.
In disciplines like medicine, a meta-analysis is
used to combine the results of independent scientific studies
on closely-related research questions (\eg establishing
the effect of a treatment for a disease), where each study
reports quantitative results (\eg a measured effect size of the
treatment) with some degree of error~\cite{Borenstein2009}. The
meta-analysis statistically derives an estimate of the
unknown common truth (\eg the true effect size of the treatment),
accounting for the errors of the individual studies. Typically, a
meta-analysis follows the random-effects model
to account for variations in study designs (\eg
different human populations)~\cite{Borenstein2009}.
Intuitively, a
random-effects-based meta-analysis estimates the true effect size as
the weighted average of the effect sizes of the individual
studies~\cite{Borenstein2009}, where the weights
are estimated via statistical methods (\eg Sidik and
Jonkman's~\cite{SidikJonkman2005}).\looseness=-1

Since the comprehensibility measurements come from
different studies with different designs (\ie with
different goals, comprehensibility interpretations and metrics, code
snippets, human subjects, \etc), a random-effects meta-analysis
is appropriate to estimate an aggregated correlation.
In our case, however, we first combine the results of the
individual correlation analyses (\ie for each metric)  for each dataset
into a single aggregated correlation per dataset, to avoid
the ``unit-of-analysis'' problem~(see \cite{harrer2021doing}, \S 3.5.2). This problem arises
in meta-analysis when there are inputs that are not independent (\ie are themselves correlated),
typically because they represent multiple measurements obtained on the same population.
Because most of our datasets include multiple metrics that were derived from
the same subjects and  snippets, and therefore, 
are related (\eg \emph{readability\_level\_ba} and \emph{readability\_level\_before}
from DS9 depend on one another), a na\"{\i}ve application of meta-analysis that treated
each metric as independent would over-weight studies with multiple metrics,
because it would ``double-count'' their statistical power (\ie multiply the statistical
power of the study by the number of metrics it contains).
We confirmed that most of the combinations of metrics within a single study
showed medium or large correlations (19/28 combinations are medium or large correlations;
of those, 13 are large), so the ``unit-of-analysis'' problem could seriously skew our results.

Dealing with the unit-of-analysis problem in meta-analyses of small
numbers of studies (as in our case) with multiple correlated metrics is an open problem
in statistical methods research. We considered the recently-proposed
correlated and hierarchical effects (CHE) model~\cite{pustejovsky2022meta}, 
but discovered that (for our data) it was highly sensitive to the choice
of the \<rho> parameter (which represents an assumption about how much variance
there is between the different metrics in each study).
Since we wanted to be conservative in our choice of statistical method,
we chose the ``brute force'' aggregation
approach suggested by~\cite{harrer2021doing}, which trades statistical
precision for simplicity and conservatism: it combines the correlation results of the various metrics in
each study into a single estimate of correlation, which guarantees that no statistical
power is derived from the presence of multiple metrics on the same population
(even if such power might be warranted). Though it also has a \<rho> parameter,
the results for our data are insensitive to the choice of \<rho>, with extremely high
and low values of \<rho> giving nearly-identical results.
All meta-analyses in \cref{sec:results} use \<rho = 0.6>.\looseness=-1

To perform the random-effects meta-analysis, we followed a standard
procedure for data preparation and
analysis~\cite{Borenstein2009}. First, we transformed
Kendall's $\tau$ values into Pearson's $\pmb{r}$
values~\cite{Walker2003}. Then, we transformed the $r$ values to be
approximately normally distributed, using Fisher's scaling.
Next, we normalized the signs of the individual metric correlations (\ie the $r$ values)
so that a negative correlation supports our hypothesis (the choice of negative is arbitrary; choosing
positive leads to the same results with the opposite sign): we multiplied by -1 the correlation value for metrics where a positive
correlation would support the hypothesis.
This strategy has been used in other disciplines when combining different metrics
whose signs have opposite interpretations, \eg in~\cite{meta-flip-signs-example}.
We used R's \textit{dmetar} package (version 0.0.9000) to aggregate the correlations of the metrics from
each study~\cite{harrer2021doing}, and the R's \textit{metafor}
package~\cite{metafor} (version 3.8-1) to run the meta-analysis and generate forest
plots to visualize the Pearson's $r$ values, their estimated
confidence intervals, the estimated weights for the aggregated
correlation, and additional meta-analysis results (\eg p-values and heterogeneity). 
\looseness=-1

To answer \textbf{RQ3}, we applied the same methodology as \textbf{RQ2}
for each individual tool's warnings (\ie no aggregation was used). We also performed a
``leave one tool out'' ablation experiment to check if any single tool
was dominating the overall meta-analysis results.
To answer \textbf{RQ4}, we repeated the same methodology
 for only the metrics in each metric category:
\emph{time}, \emph{correctness}, \emph{rating}, and
\emph{physiological}---\ie we performed four
meta-analyses, one for each metric group.

While we provide the $p$-values of all of these statistical analyses, we
emphasize that they
should be interpreted with caution given the relatively small sample sizes
(and, for \textbf{RQ1}, that fact that 20 metrics are considered). For
example, DS2 only contains 12 snippets, which means only 12 data
points were used for correlation for its metrics.
We also used a meta-analysis (\textbf{RQ2-RQ4}) because interpreting the individual metric results (\textbf{RQ1}) to draw general conclusions for our hypothesis can be misleading~\cite{Borenstein2009-vote-counting}. 
Our meta-analysis also obviates the need for statistical
correction to avoid multiple comparisons, such as Holm-Bonferroni's~\cite{holm1979simple}:
the meta-analysis aggregates all of the results and informs us of the overall trend.
We use the same interpretation guidelines for
Pearson's $r$ values that we used for Kendall's $\tau$: \emph{small} when $0.1 \leq |r| < 0.3$, \etc~\cite{Cohen:2002,Peitek:ICSE21,Scalabrino:TSE19}.

\section{Study Results and Discussion}
\label{sec:results}

We present and discuss the results of our study in this section.
Scripts and data that generate these results %
are available in our replication package~\cite{repl-pack}. 

\subsection{RQ1: Individual Correlation Results}
\label{sec:answer-rq1}

\newcommand{\rqOneLine}[5]{\small{\emph{#1}} & #2 & #3 & #4 & #5 }

\begin{table}
  \caption{Correlation results based on Kendall's $\pmb{\tau}$
    (\underline{K.'s~$\pmb{\tau}$}) for each dataset (\textbf{DS}) and
    \underline{Metric}.
    A metric falls into one \underline{Type}:
    \underline{C}orrectness, \underline{T}ime, \underline{R}ating, \&
    \underline{P}hysiological.
    The expected correlation direction 
    (\underline{Exp. Cor.}), if our hypothesis is correct, is either Positive or Negative.
    We assess $\pmb{\tau}$'s direction/strength, compared to the expected correlation
    (\underline{Exp?}): \textbf{`-'} means \textit{no} correlation,
    `Y/y' means expected and measured correlations match (thus
    supporting our hypothesis), and `N/n' means they do not
    match. Capital letters in darker colors (\colorbox{green!40} {\textbf{Y}}/\colorbox{red!40} {\textbf{N}}) mean a
     \textit{medium} or higher correlation. Lowercase letters and lighter colors (\colorbox{green!20} {\textbf{y}}/\colorbox{red!20} {\textbf{n}}) mean a
    \textit{small} correlation. $\pmb{\tau}$'s significance is tested at the
    $\pmb{p<0.05}$ (\textbf{*}) \& $\pmb{p<0.01}$ levels~(\textbf{**}).  }
\label{tab:correlation-analysis-summary}
\centering
\setlength\tabcolsep{3.5pt}
\begin{tabular}{clcccc}
  \hline
   \textbf{DS} & \textbf{Metric} & \textbf{Type} & \textbf{Exp. Cor.} & \textbf{K.'s $\pmb{\tau}$} & \textbf{Exp?} \\
\hline
\multirow{3}{*}{1} & \rqOneLine{correct\_output\_rating}{C}{Negative}{-0.34*}{\colorbox{green!40} {\textbf{Y}}} \\
& \rqOneLine{output\_difficulty}{R}{Negative}{-0.43**}{\colorbox{green!40} {\textbf{Y}}} \\
& \rqOneLine{time\_to\_give\_output}{T}{Positive}{0.41**}{\colorbox{green!40} {\textbf{Y}}} \\
\hline
\multirow{4}{*}{2} & \rqOneLine{brain\_deact\_31ant}{P}{Negative}{-0.31}{\colorbox{green!40} {\textbf{Y}}} \\
& \rqOneLine{brain\_deact\_31post}{P}{Negative}{-0.45}{\colorbox{green!40} {\textbf{Y}}} \\
& \rqOneLine{brain\_deact\_32}{P}{Negative}{-0.38}{\colorbox{green!40} {\textbf{Y}}} \\
& \rqOneLine{time\_to\_understand}{T}{Positive}{0.14}{\colorbox{green!20} {\textbf{y}}} \\
\hline
\multirow{1}{*}{3} & \rqOneLine{readability\_level}{R}{Negative}{-0.17*}{\colorbox{green!20} {\textbf{y}}} \\
\hline
\multirow{3}{*}{6} & \rqOneLine{binary\_understand}{R}{Negative}{\rev{0.01}}{-} \\
& \rqOneLine{correct\_verif\_questions}{C}{Negative}{\rev{0.02}}{-} \\
& \rqOneLine{time\_to\_understand}{T}{Positive}{0.05}{-} \\
\hline
\multirow{4}{*}{9} & \rqOneLine{gap\_accuracy}{C}{Negative}{\rev{-0.34}}{\colorbox{green!40} {\textbf{Y}}} \\
& \rqOneLine{readability\_level\_ba}{R}{Negative}{\rev{0.08}}{\rev{-}} \\
& \rqOneLine{readability\_level\_before}{R}{Negative}{\rev{0.13}}{\colorbox{red!20} {\textbf{n}}} \\
& \rqOneLine{time\_to\_read\_complete}{T}{Positive}{\rev{-0.23}}{\colorbox{red!20} {\rev{\textbf{n}}}} \\
\hline
\multirow{5}{*}{F} & \rqOneLine{brain\_deact\_31}{P}{Negative}{-0.18}{\colorbox{green!20} {\textbf{y}}} \\
& \rqOneLine{brain\_deact\_32}{P}{Negative}{-0.18}{\colorbox{green!20} {\textbf{y}}} \\
& \rqOneLine{complexity\_level}{R}{Positive}{0.35}{\colorbox{green!40} {\textbf{Y}}} \\
& \rqOneLine{perc\_correct\_out}{C}{Negative}{-0.16}{\colorbox{green!20} {\textbf{y}}} \\
& \rqOneLine{time\_to\_understand}{T}{Positive}{-0.13}{\colorbox{red!20} {\textbf{n}}} \\
\hline
\end{tabular}
\posttablecaption
\end{table}

\Cref{tab:correlation-analysis-summary} summarizes the results of
each metric's
correlation (based on Kendall's~$\tau$) with the total number of warnings from all tools.
We provide descriptive statistics about these results, but we emphasize that
these results should be interpreted with caution: ``vote-counting'' (\eg checking
the number of statistically-significant metrics in each direction) can lead
to misleading conclusions~\cite{Borenstein2009-vote-counting}. We avoid directly drawing conclusions from these results and instead, we investigate the aggregated trend of all metrics with a meta-analysis
in \cref{sec:answer-rq2}.

\Cref{tab:correlation-analysis-summary} shows that 
for 13 of the 20 (65\%) metrics,
the direction of the correlation
supports our hypothesis.
For \rev{4} metrics, there is \emph{no}
correlation, %
and for the remaining {3} metrics, the correlation is in
the opposite direction than expected.
\Cref{tab:correlation-analysis-summary} indicates the strength of
the correlation in the rightmost column.
Of the metrics where we found
a \textit{medium} or higher correlation, 8/8 are in the direction that supports
our hypothesis. For the other 5 metrics that support our hypothesis
and the \rev{3} metrics that do not, their correlation is \textit{small}.
While we cannot directly draw conclusions from these results about the overall trend,
they are suggestive. We examine the aggregate trend rigorously with a meta-analysis
in \cref{sec:answer-rq2}.

With regard to metric categories, 3/4 correctness (\textbf{C}) metrics, 3/6
rating (\textbf{R}) metrics, 2/5 time (\textbf{T}) metrics, and 5/5 physiological (\textbf{P}) metrics
correlate with verifiability.
All \rev{3} metrics that anti-correlate with verifiability
are concentrated in the rating
and time categories.
These two metric categories are the most subjective:
ratings are opinions, and some time metrics require the human subjects to
signal the experimenter when they complete the task.
These results suggest that there may be a relationship between metric categories
and the correlation with verifiability; we further investigate the differences between metric categories
in \cref{sec:answer-rq4}.

\subsection{RQ2: Aggregate Correlation Results}
\label{sec:answer-rq2}

\begin{figure}
  \centering
  \includegraphics[width=\linewidth]{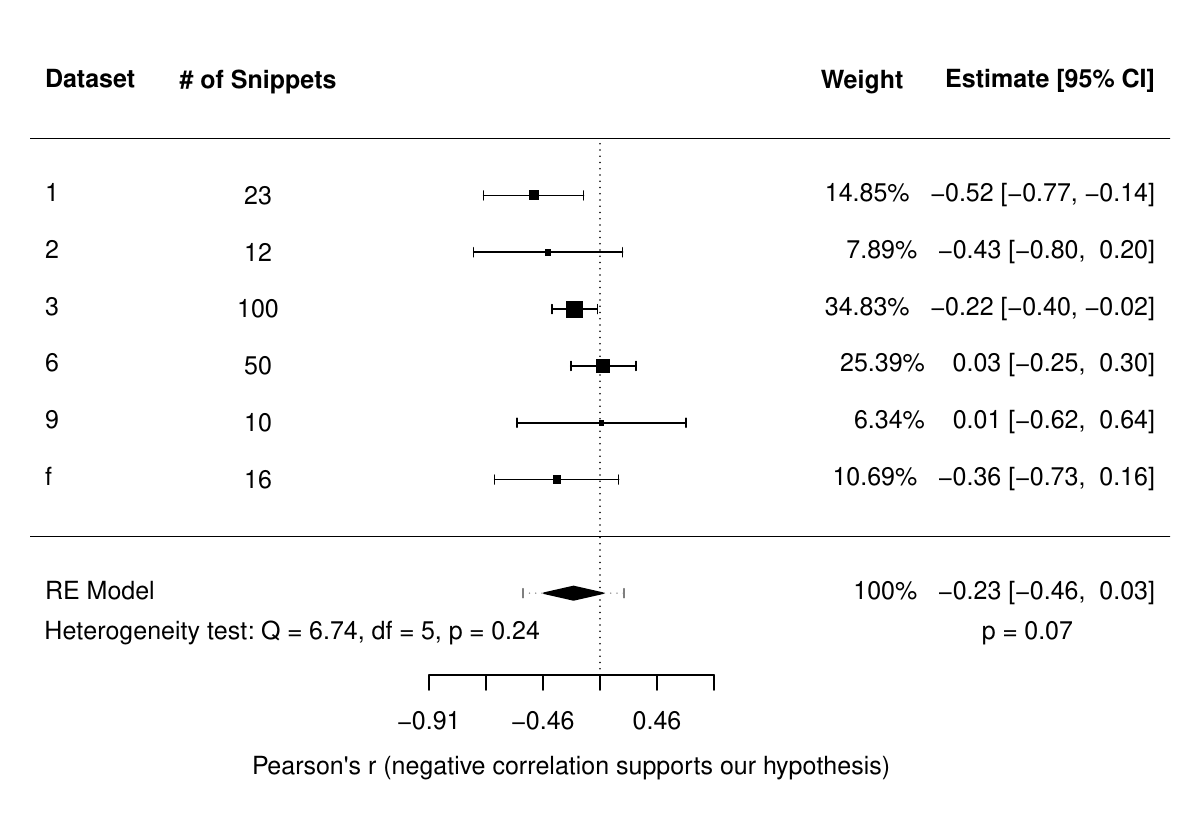}
  \caption{\os{To make this Figure 1 pretty (we need to make the text more readable)}Results of the random-effects meta-analysis of the metrics in \cref{tab:correlation-analysis-summary},
    after aggregating the results by dataset to avoid the unit-of-analysis problem (\cref{sec:correlation-methods}).}
  \label{fig:meta-analysis-summary}
\end{figure}

Because direct interpretation of \cref{tab:correlation-analysis-summary}
is difficult due to the different sample sizes of the various studies,
we performed a meta-analysis to understand the overall trend (see \cref{fig:meta-analysis-summary}).
As noted in \cref{sec:correlation-methods}, the results are presented by dataset rather than
by metric to avoid unit-of-analysis errors~\cite{harrer2021doing}.

The forest plot in \cref{fig:meta-analysis-summary} displays the observed correlation
(Person's \textbf{$\pmb{r}$ value}, obtained from
the Kendall's $\tau$ value as described in \cref{sec:correlation-methods}) and
the 95\% confidence interval (\textbf{Estimate [95\% CI]}),
as well as the estimated weight for each dataset (\textbf{Weight}).
This information is shown numerically and graphically
in \cref{fig:meta-analysis-summary}.
Each box's size is the dataset's estimated weight
(a larger box size means a larger weight), and the box's middle point
represents the correlation with respect to the dashed vertical line at zero.
There is a negative correlation if the box is to the
left of the vertical line; positive if it is to the right;
all metrics have been normalized so that the expected correlation
is negative (that is, a negative correlation supports our hypothesis).
The horizontal lines visualize the confidence
intervals for each dataset. 
At the bottom, the plot shows the aggregated correlation
(on the right) and related information
calculated by the meta-analysis.
The diamond at the bottom of the plot visualizes the aggregated correlation;
the width of the diamond represents the confidence interval.

\Cref{fig:meta-analysis-summary} shows a small aggregated correlation
supporting an affirmative answer to \textbf{RQ2} ($r=-0.23$, with a 95\% CI
that contains negligible, small, medium correlations: $r=-0.46$ to $r=0.03$, $p=0.07$).
We interpret these results overall as support for the hypothesis that
tool-based verifiability and humans' ability to understand code
are correlated to some extent. 
\looseness=-1

The heterogeneity of the considered studies is non-negligible ($I^{2} = (Q  - df) / Q =$ \rev{$(6.74 - 5) / 6.74 = 25.8\%$} -- not shown in \cref{fig:meta-analysis-summary}), indicating
that  \rev{25.8\%} of the correlation variation (\ie variance) we observe is due to the studies measuring different
factors rather than due to chance. This result validates our choice of a random-effects model for the meta-analysis. 

The plots in \cref{fig:meta-analysis-summary} show wide confidence
intervals for all the datasets except DS 3 and DS6, which indicates relatively high
variability in the correlations. This indicates that most of these
studies were under-powered for our purpose: the number of snippets
considered was not high enough to give the meta-analysis much confidence
in the correlation results. The meta-analysis correspondingly gives the
largest weights to the two datasets with the most snippets: about 35\% weight
to DS3 (with 100 snippets) and about 25\% weight to DS6 (with 50 snippets).
Future work should explore running understandability experiments with larger
numbers of snippets, which would enable us to gain further confidence in our
results.

\subsection{RQ3: Correlation Results By Tool}
\label{sec:answer-rq3}

To answer \textbf{RQ3}, we repeated the analyses used to answer \textbf{RQ1}
and \textbf{RQ2}
independently for each tool (\ie no
warning aggregation). We also repeated the analysis in a ``leave-one-out''
ablation experiment. We report only the summary results for each tool (\ie the
results of the meta-analyses) for space reasons; forest plots similar to
\cref{fig:meta-analysis-summary} as well as the individual correlation results on
each tool+metric combination are available in our replication package~\cite{repl-pack}.

Repeating the meta-analysis on only the warnings
produced by each tool individually gave similar results to the meta-analysis
in \cref{fig:meta-analysis-summary}, \rev{except for Infer}. 
The Checker Framework results
supports our hypothesis more strongly than the overall meta-analysis ($r=-0.26$, 95\% CI of $[-0.44, -0.06]$,
$p = 0.02$). The results of OpenJML and JaTyC support the hypothesis more weakly than the overall
results ($r=-0.12$, with a 95\% CI of $[-0.29, 0.07]$, $p = 0.16$ for OpenJML and $r=-0.17$ with a 95\% CI of
$[-0.39, 0.08]$, $p=0.14$ for JaTyC). \rev{Infer has too few warnings to draw meaningful conclusions from its
results ($r=-0.09$ with a 95\% CI of $[-0.94, 0.91]$, $p=0.60$).} 

From these results, we conclude that, while some tools support the hypothesis
less strongly than the overall meta-analysis, all the tools \rev{but Infer} show the same trend. %
These results support the overall meta-analysis results (\textbf{RQ2}): the correlation
measured for \rev{3 of the 4} studied tools suggests that the correlation between verifiability and understandability indeed exists (in small magnitude); no tool shows a markedly different trend \rev{except Infer, whose trend is not meaningful due to its small warning count}.

We were also concerned that a single tool might be dominating the overall results.
To mitigate this threat, we
performed an ablation study by
repeating the meta-analysis on warning data aggregated from each combination of three tools (\ie excluding the warnings of one tool only). Overall, the results are extremely similar for each combination of tools to the overall results---the results without
Infer are in fact nearly identical---with $r$ values ranging from $-0.23$ to $-0.20$; CI lowerbounds ranging from $-0.46$ to $-0.37$,
and CI upperbounds ranging from $-0.01$ to $0.06$; and $p$ values from $0.04$ to $0.10$.
We conclude from this ablation experiment that no single tool dominates
the \textbf{RQ2} results.

Taken together, the results in this section show that the correlation found for \textbf{RQ2}
is not entirely driven by any tool: the overall results remain similar (if slightly weaker) for every tool individually \rev{except Infer}
and for each combination of three tools (\ie without each tool).
We interpret these results to mean that the correlation exists 
regardless of the specific verifier in use---meaning that our results apply to verification in general.

\subsection{RQ4: Correlation Results by Metric Type}
\label{sec:answer-rq4}

In \cref{sec:answer-rq1}, we observed that the correctness and physiological
metric categories appeared to support our hypothesis more strongly than the
rating and time categories. To test this observation, we repeated
our meta-analyses for each of the four metric categories.

The results refute the idea that these categories are a major influence
on the results. The correctness, rating, and time metrics show
 overall results similar to \cref{fig:meta-analysis-summary}, but with %
wider confidence intervals: \rev{$r=-0.28$} with 95\% CI of $[-0.70, 0.27]$, \rev{$p=0.20$} for correctness;
$r=-0.25$ with 95\% CI of \rev{$[-0.54, 0.09]$, $p=0.11$} for rating;
and \rev{$r=-0.22$ with 95\% CI of $[-0.58, 0.21]$, $p=0.23$} for time. The results for the physiological metrics show that they have a minimal impact: 
\rev{$r=-0.32$} but with a huge 95\% CI of $[-1.00, 0.98]$.
These results, especially for the physiological metrics, are likely due to the smaller sample
sizes created by considering only one metric type; \eg there are physiological metrics
in only two datasets (DS2 and DSF) with 28 total snippets between them. The dataset with the most weight
in the overall results (DS3) only has rating metrics, which reduces the meta-analysis' confidence in
the other types. Finally, the heterogeneity for the three metric categories with
useful results (\ie not physiological) is higher than in the overall results (with $I^{2}$ of \rev{57\%, 49\%, and 50\%} for
correctness, rating, and time metrics, respectively).

%

%
%
%
%
%

\begin{comment}
  Interpretation of I^2
 25%
 50%
 75%
\end{comment}

%

%
%
%
%
%
%
%
%
%
%
%
%
%
%
%
%

%
%
%
%
%
%
%
%
%
%
%
%
%

%
%
%
%
%
%
%
%
%

\subsection{Robustness Experiments}

We ran additional experiments to probe the robustness of
the findings for the RQs and mitigate
some threats to validity.

\subsubsection{Handling Code Comments in Dataset 9}
\label{sec:dataset9-comments}

DS9's original study had 3 versions of each of its 10 snippets, with three types of code comments: ``good'', ``bad'',
and no comments~\cite{Borstler:TSE16}.  The results presented elsewhere in this section used the ``No comments'' (NC)
version of DS9, because none of the four verifiers use comments
as part of their logic. However, this choice might be source of possible bias,
so we analyzed how the
correlation results would change if we had used the ``Good comments'' (GC)
or ``Bad comments'' (BC) versions of the dataset. Note that because none of
the verifiers take comments into account, their warnings
are exactly the same---the only differences are in the comprehensibility measurements.
\Cref{tab:dataset9} shows how the correlation results
differ for the three versions of DS9. A significant difference
is observed in the two readability metrics: when the comments are bad, these
metrics are anti-correlated with verifiability: that is, humans
rated the snippets on which the tools issued more warnings as \emph{more
readable}. We see a similar phenomenon for the time metrics,
but it occurs only for the good (rather than bad) comments.
To explain this phenomenon, we compared the distribution of the metrics across comment categories and analyzed the scatter plots of the data used for correlation. Our analysis revealed that such disparity in correlation stems from a combination of (1) outliers found in the human measurements (likely due to data collection imprecisions in the original study~\cite{Borstler:TSE16}) and (2) the low number of  data points in DS9. For example, we found that bad comment code
was rated more readable by a few participants than code with no comments, even though the snippets were semantically the same. These unusual measurements led to outliers that had a considerable impact on the correlation results across comment categories (because of the small number of data points).
The effect of this phenomenon on the overall results is low, because DS9 is given
very low weight (6.34\%, lowest among all datasets) by the meta-analysis due to its small sample size.

\newcommand{\dsNineLine}[5]{ \small{\emph{#1}} & #2 & #3 & #4 & #5}

\begin{table}
  \caption{Correlation results (\textbf{Kendall's $\pmb{\tau}$}) on different versions of DS9: "No" (\textbf{NC}), "Bad" (\textbf{BC}), and "Good Comments" (\textbf{GC}). A \textbf{**} indicates statistical significance at the $\pmb{p<0.01}$ level.}
  \begin{tabularx}{\columnwidth}{Xcccc}
    \hline
    \textbf{Metric} & \textbf{Exp. Cor.} & \textbf{NC} & \textbf{BC} & \textbf{GC} \\
    \hline
    \dsNineLine{gap\_accuracy}{Negative}{\rev{-0.34}}{\rev{-0.18}}{\rev{-0.34}} \\
    \dsNineLine{readability\_level\_ba}{Negative}{\rev{0.08}}{\rev{0.44}}{\rev{-0.18}} \\
    \dsNineLine{readability\_level\_before}{Negative}{\rev{0.13}}{\rev{0.42}}{\rev{-0.05}} \\
    \dsNineLine{time\_to\_read\_complete}{Positive}{\rev{-0.23}}{\rev{-0.39}}{\rev{-0.75**}} \\
    \hline
  \end{tabularx}
  \label{tab:dataset9}
  \posttablecaption
\end{table}

\subsubsection{Handling OpenJML Timeouts}
\label{sec:openjml-timeouts}

\newcommand{\openjmlLine}[4]{\small{\emph{#1}} & #2 & #3 & #4}

\begin{table}
  \caption{Correlation results (\textbf{Kendall's $\pmb{\tau}$}) for OpenJML, for each timeout-handling approach: (1) \underline{ignore} timeouts; (2) \underline{under}-estimate the warnings hidden by timeouts; (3) \underline{over}-estimate the warnings hidden by timeouts. $\pmb{\tau}$'s significance is tested at the
  $\pmb{p<0.05}$ (\textbf{*}) and $\pmb{p<0.01}$ levels~(\textbf{**}).
  }
  \begin{tabular}{clccc}
    \hline
    & & \multicolumn{3}{c}{\textbf{Approach}} \\ \cline{3-5}
    \textbf{DS} & \textbf{Metric} & \textbf{1: Ignore} & \textbf{2: Under} & \textbf{3: Over} \\
    \hline
    \multirow{1}{*}{3} & \openjmlLine{readability\_level}{-0.20**}{-0.23**}{-0.17*} \\
    \hline
    \multirow{3}{*}{6} & \openjmlLine{binary\_understand}{-0.07}{-0.07}{0.00} \\
    & \openjmlLine{correct\_verif}{-0.06}{-0.06}{0.00} \\
    & \openjmlLine{time\_to\_understand}{0.11}{0.11}{0.05} \\
    \hline
  \end{tabular}
  \label{tab:openjml-timeouts}
  \posttablecaption
\end{table}

OpenJML uses an SMT solver under the hood. Though modern SMT solvers
return results quickly for most queries using sophisticated heuristics,
some queries do lead to exponential run time, making
it necessary to set a timeout when analyzing a collection of snippets.
We used a 60 minute timeout, which led to
2/50 snippets in DS6
and 39/100 snippets in DS3 timing out
(and zero in the other datasets).
We considered three approaches in our correlation analysis to handle
timeouts: (1) ignore snippets containing timeouts entirely,
(2) count each timeout as zero warnings (but do count any other warnings
issued in the snippet before timing out), or
(3) count each snippet that timed out as the maximum
warning count in the dataset.
All the results for RQ1-RQ4 were produced by following approach 3.
The reason we chose approach 3 over approach 2 is that timeouts
typically occur on the most complicated SMT queries, which might hide
many warnings. Therefore, approach 2 \emph{underestimates} the warning count
that a no-timeout run of OpenJML would encounter, while approach 3
\emph{overestimates} the warning count in a no-timeout run. We re-ran
the correlation analysis under all three conditions. The results
are in \cref{tab:openjml-timeouts} and do not show any significant
differences between the strategies for timeouts---the overall direction and
strength of the correlations are similar, and the absolute size of the differences
is small, meaning that the impact on the meta-analysis is negligible.

\subsection{Results Discussion and Implications}
\label{sec:implications}

\rev{
}

\subsubsection{Program Semantics and Understandability.} 
\rev{Verification warning counts (indirectly) encode \textit{program semantics} rather than syntactic properties of the code. Verification tools are trying to prove semantic
properties: checking syntactic properties is decidable, so it is not the target of
verifiers (which find approximate solutions to undecidable, semantic problems, \eg using SMT solvers). Our results suggest that there
might be complexity caused by semantics, and verifiers are well suited to reasoning about that kind of complexity. Previous work using decidable syntactic metrics for complexity~\cite{Nunez-Varela:JSS17,Curtis:TSE79,Zuse:IWCP'93,Sneed:JSMRP95,Ajami:EMSE19,Jbara:EMSE17,Scalabrino:TSE19,Tahir:ICSM12} certainly
could not capture semantics (since any non-trival semantic property of a program is undecidable~\cite{rice1953classes})---see \cref{sec:related-work}.
\looseness=-1

On one side, the measured correlation between verifiability and
understandability increases our confidence that there is a semantic component to human
code understanding. On the other side, the small correlation we measured indicates that there are other factors to code understanding beyond \textit{just} program semantics.
Neither of these conclusions are particularly surprising, but program understandability
research has so far mostly focused on the non-semantic components (such as variable
names or syntactic metrics---see \cref{sec:related-work}). Our work motivates the need for future studies that investigate the
\emph{semantic} component of code understanding; in particular, the specific semantic factors make code simple or complex, and how they impact understandability. Fortunately, our work also offers a path
forward: the verification community has already built many tools that attempt to verify
semantic properties (\ie verifiers), which gives us an opportunity to leverage those
existing tools to improve our understanding of code complexity and understandability.
\looseness=-1}

\subsubsection{Incorporating Verifiability into Comprehensibility Models.} 
\rev{Most prior attempts to design automated metrics or models that measure or predict code understandability
have used \emph{syntactic} features that do not account for
program semantics~\cite{Nunez-Varela:JSS17,Curtis:TSE79,Zuse:IWCP'93,Sneed:JSMRP95,Ajami:EMSE19,Jbara:EMSE17,Scalabrino:TSE19,Tahir:ICSM12}. Rather, they used syntactic features such as code branching, vocabulary size, and executions paths, among other proxies that attempt to capture code complexity (see \cref{sec:related-work}). Many of these features have shown to be poor predictors of code understandability~\cite{Ajami:EMSE19,Jbara:EMSE17,Scalabrino:TSE19,Kaner2004,Feigenspan:ESEM11}. We believe one of the reasons for this is because they do not 
capture complexity arising from program semantics. 
Based on the link we found between verifiability and comprehensibility, we hypothesize that semantic code properties would lead to more accurate models of understandability.
}

\rev{Future work should validate this hypothesis by incorporating verifiability into models that predict human-based
comprehensibility~\cite{Raymond:TSE10,Scalabrino:TSE19,Trockman:MSR18} and measuring its impact on prediction performance.
If the link between verifiability and comprehensibility exists (as our
results suggest), verifiability information should complement the
syntactic features of these models. Verifiability can be
captured by adapting existing verification tools or by leveraging tool warning data. For example, we could provide the number of warnings a tool produces on a snippet as
an input feature to these models. 
\looseness=-1}

\subsubsection{Reducing False Positives to Increase Code Comprehensibility.}
\rev{Developers could use the warning count of verifiers to know when code might be complex, \ie when it might need to be refactored to reduce \textit{accidental} complexity. While coding, the developer can monitor the warning count of verifiers on a code snippet (\eg a method they are writing or updating), knowing the code is correct. If this count increases, they could assess potential complex parts of the method and come up with changes to the method that would be semantically equivalent (\eg replacing recursion, which is traditionally hard for verifiers to
reason about, with a loop). This auxiliary benefit of using verification tools has not
been studied in the literature, and might represent an opportunity to make verifiers more
appealing to everyday developers.}

\rev{This usage scenario poses a research opportunity too: what if we could automatically determine and suggest to the developer a semantically-equivalent refactoring %
that is easier to verify? Such a refactoring would change the code to perform the same task, but would cause a verifier to issue fewer warnings. The measured correlation between verifiability and understandability would mean, more often than not, that applying such a refactoring might make the code easier to understand. Since it is unclear if such refactoring is possible, more research should be conducted. However, if it is possible and the developer is aware of the correlation, we anticipate they would be more willing to (1) use verifiers in their everyday coding tasks, and (2) accept the refactoring suggestion. One possible issue with this approach
is that our correlation includes warnings caused by missing specifications; there is large existing
literature on specification inference (\eg~\cite{houdini,damas1982principal,vakilian2015cascade}) that could be leveraged to focus only on false positives when specifications are explicit.} 

\rev{
\subsubsection{Code Verifiability \textit{vs.} Understandability \textit{vs.} Complexity}
    Our study found a correlation between code understandability and verifiability, yet it did not find whether one of the two causes the other (\ie correlation does not imply causation). Further research is needed
    to determine whether one causes the other, or whether there are other factors that
    cause both. However, based on our results and discussion, we hypothesize that \emph{code complexity}
    causes both humans and verification tools to struggle to understand code.
    Future studies should investigate this and other possible causes.
}

\section{Limitations and Threats to Validity}
\label{sec:threats}

\os{add somewhere here the limitation that we cannot double check that simpler code snippets selected by prior studies indeed have fewer false positive. This is because the prior studies did not control for complexity. They selected snippets with different varieties of complexity (how do we know this?)}

Our study shows a \emph{correlation} between verifiability and
understandability, but do not show one \emph{causes} the
other. So, our results must be interpreted
carefully: further work is needed to determine causality.\looseness=-1

Regarding threats to external validity, the correlation
we found may not generalize beyond the specific conditions of our study.
The snippets are all Java code, so the results
may not generalize to other languages. We only used
a few verifiers, as we were limited by parcity
of practical tools that can analyze the snippets. While
limitations or bugs in individual tools could skew our results,
we mitigated this threat by re-running the experiments
individually for each tool 
and with an ablation experiment (\cref{sec:answer-rq3}), which demonstrated
that no single tool dominates the results.
The snippets are small compared to
full programs; the comprehensibility of larger programs may differ.
Further, 3/6 datasets are
snippets from introductory CS courses rather than real-life programs, but
this is mitigated by the other three datasets of open-source snippets.

Another threat is that subjects in the prior studies were mostly students.
Only DS6 used professional software engineers (and only 13/63 participants---the
other 50 were students),
so our results may not apply to more experienced programmers. Future work
should conduct understandability studies with professional engineers.

Beyond the datasets and tools, there are threats to
internal and construct validity. 
\rev{We assumed the snippets are correct as written, and that each verifier
  warning therefore represents either a false positive or a specification that a human would need
  to write to verify the code.}
The presence of a bug
would make a snippet seem ``harder to verify'' in our analysis
(because every verifier would warn about it), even if the snippet is
easy for humans to understand, skewing the results.
\rev{We mitigated this threat by manually examining a representative subset of the warnings
  as described in \cref{sec:tool-execution}; %
  we did not observe any bugs in the snippets.}
%
%
%

\begin{comment}
\begin{itemize}
\item  Tools (the actual tools we selected and low \# of tools)
\item  Datasets (simple snippets, only five datasets, we did alternations to some of the snippets to make them compilable?, and comprehensibility measurements for students mostly)
\item  Java (other languages is future work)
\item  Average of comprehensibility measurements (for Likert scale measurements)
\item  Dataset 9 (comments)
\item  We are doing correlation, which is not causation
\item  Some warnings might be true positives, if the snippets are buggy. \todo{Martin could spot-check some of the warnings to mitigate this threat.}
\end{itemize}
\end{comment}

%

\section{Related Work}
\label{sec:related-work}

\textbf{Code complexity metrics.} Researchers have proposed many metrics
for code
complexity~\cite{Nunez-Varela:JSS17,Curtis:TSE79,Zuse:IWCP'93,Sneed:JSMRP95,Ajami:EMSE19,Jbara:EMSE17,Scalabrino:TSE19},
though the concept is not easy to define due to
different interpretations~\cite{Antinyan:EMSE17,Antinyan:IEEE20}. Most
metrics rely on simple, syntactic properties
such as code size or branching paths~\cite{Peitek:ICSE21,Nunez-Varela:JSS17}.
These metrics are used to detect complex code
so developers can simplify it during
software evolution~\cite{Garcia-Munoz2016,Peitek:ICSE21,Ammerlaan:SANER15}.
The motivation is that complex code is harder to understand~\cite{Scalabrino:TSE19,Ajami:EMSE19},
which may have important repercussions on developer effort
and software quality (\eg bugs introduced due to misunderstood code).
Our correlation results imply that code that is easier to
verify might also be simple and easier to
understand by humans; we believe the underlying mechanism
might be that simple code fits into the expected code patterns
of a verification technique. Our results also suggest that a complexity
metric that aims to capture human understandability should consider
not only syntactic information about the code, but also its semantics.
\textbf{Empirical validation of complexity metrics.} %
Scalabrino \etal~\cite{Scalabrino:TSE19} collected
code understandability measurements from developers and students
on open-source code. %
They correlated their measurements with 121 syntactic complexity
metrics (\eg cyclomatic complexity, LOC,
\etc) and developer-related properties
(\eg code author's experience and background).
They found small correlations for only a few metrics,
but a model trained on combinations of metrics performed better.
Another study found similar results~\cite{Trockman:MSR18}.\looseness=-1

Researchers have explored the limitations of classical complexity
metrics~\cite{Ajami:EMSE19,Jbara:EMSE17,Scalabrino:TSE19,Kaner2004,Feigenspan:ESEM11}.
For example, Ajami \etal~\cite{Ajami:EMSE19} found that different code
constructs (\eg~\<if>s vs. \<for> loops) have different
effects on how developers comprehend code, implying that metrics
such as cyclomatic complexity, which weights code constructs equally,
fail to capture understandability~\cite{Peitek:ICSE21}.
Recent work has proposed new metrics
such as Cognitive Complexity
(COG)~\cite{Campbell:TechDebt18,Saborido:IEEE22}, which assigns
different weights to different
code constructs.
Muñoz \etal~\cite{MunozBaron:ESEM20}
conducted a correlation meta-analysis between COG and human understandability.
They found that time and rating metrics
have a modest correlation with COG, while
correctness and physiological metrics have no correlation.
They did not take into account the unit-of-analysis problem
in their meta-analyses.

We extend prior work with empirical evidence of the correlation between verifiability and human understandability.
To the best of our knowledge, we are the first to investigate this empirically.\looseness=-1
\textbf{Studying code understandability.}
Researchers have studied code understandability
and factors affecting it
via controlled experiments and user
studies~\cite{Siegmund:ICSE14,Peitek:TSE18,Peitek:ICSE21,Borstler:TSE16,Ajami:EMSE19,Jbara:EMSE17,Johnson:ICSME19,Gopstein:FSE20}.
Precisely defining understandability is difficult,
so some studies~\cite{Raymond:TSE10,Borstler:TSE16,MunozBaron:ESEM20,Piantadosi:EMSE20,Oliveira:ICSME20} use it interchangeably with
readability
(a different, yet related concept).
Measurements include the time to
read, understand, or complete code; the correctness of
output given by the participants; perceived code complexity, readability
or understandability; and
(recently) physiological measures from
fMRI scanners~\cite{Siegmund:ICSE14,Peitek:TSE18,Peitek:ICSE21},
biometrics
sensors~\cite{Fritz:ICSE14,Fucci:ICPC19,Yeh:FIE17},
or eye-tracking devices~\cite{Fritz:ICSE14,Turner:ETRA14,Binkley:EMSE13,Abbad-Andaloussi:ICPC22}.
Our study utilizes these human-based measurements of understandability to assess their correlation with verifiability.\looseness=-1

Factors that affect understandability include: code
constructs~\cite{Ajami:EMSE19,Johnson:ICSME19} and patterns \cite{Jbara:EMSE17,Langhout:ICPC21,Borstler:TSE16},
identifier quality and style~\cite{Wiese:ICSE19,Siegmund:FSE17}, %
comments~\cite{Borstler:TSE16}, information gathering tasks~\cite{LaToza:FSE07,Siegmund:ICSE14,Siegmund:FSE17,Binkley:EMSE13},
comprehension tools~\cite{Storey:SCP00}, code reading
behavior~\cite{Peitek:ICPC20,Abid:ICSE19,Siegmund:SANER16},
authorship~\cite{Fritz:ICSE10}, high-level comprehension strategies~\cite{Siegmund:SANER16}, programmer
experience~\cite{Wiese:ICSE19,Xia:TSE18}, and 
the use of complexity metrics~\cite{Wyrich:ICSE21}.
Our work investigates a new factor impacting understandability: code verifiability. %
Our results suggest there is a correlation between these variables,
yet future studies are needed to assess causality.\looseness=-1

\textbf{Studies of verification and static analysis tools.}
A study conducted
to evaluate a code readability model~\cite{Raymond:TSE10} is closely related to ours.
The model was found to correlate moderately with snippets
on which FindBugs~\cite{AyewahHMPP2008} issued warnings. Unlike
the tools in our study, FindBugs is \emph{not} a verification
tool (it uses heuristics to flag possibly-buggy code).
We correlated verifiability with human understandability;
the earlier study correlated FindBugs warnings with an automated
readability \emph{model} trained on human judgments.\looseness=-1

Though verification and static analysis tools are becoming more
common in industry~\cite{Beller:SANER16,RutarAF2004},
studies of their use and the challenges developers face in deploying
them~\cite{Vassallo:EMSE20,Beller:SANER16,Nachtigall:ISSTA22,Smith:SOUPS20,Mansoor:SANER22}
suggest that false positives remain a problem
in practice~\cite{Johnson:ICSE13,Nachtigall:ISSTA22}.
Our work gives a new perspective the problem of false positives.
We have shown that the presence of false positives from verifiers
correlates with more difficult-to-understand code. We hope that this perspective encourages developers to view false
positives as opportunities to improve their code
rather than as barriers to finding defects~\cite{Sadowski:ICSE15}.

\section{Conclusions and Future Work}

Our empirical study on the correlation between tool-based verifiability
and human-based metrics of code understanding suggests there \emph{is}
a connection between whether a tool can verify a code snippet and
how easy it is for a human to understand.
Though our results are suggestive, our meta-analysis shows that
extant studies on human code understandability lack sufficient power
to enable us to draw a stronger conclusion, so more studies of understandability
(preferably including many more snippets of code) are needed.
Further, our work has shown only a correlation: establishing a causal link
between verifiability and understandability---perhaps through a mutual cause,
such as  complexity---remains future work.

Verifiability is a promising
alternative that complements traditional code complexity metrics, and future work could
combine measures of tool-based verifiability with modern complexity
metrics such as cognitive complexity that seem to capture different aspects
of human understandability into a unified, automatic model. Our results are also
promising support for the prospect of increased adoption of verifiers:
our results offer a new perspective on the classic problem of false positives,
since they suggest that false positives from verifiers are opportunities
to identify potentially more complex code and make it more understandable by humans.\looseness=-1

\section*{Acknowledgements}

We thank the anonymous reviewers and Michael D. Ernst for comments that helped
to improve this paper. Ji Meng Loh provided invaluable advice about our statistical
approach.

\bibliographystyle{ACM-Reference-Format}
\balance
%%% -*-BibTeX-*-
%%% Do NOT edit. File created by BibTeX with style
%%% ACM-Reference-Format-Journals [18-Jan-2012].

%
%

\end{document}